# Complex Networks on a Rock Joint

H.O.Ghaffari, M. Fall, E. Evgin
*Civil Engineering Department, University of Ottawa, Ottawa, Ontario, Canada*

M.Sharifzadeh
*Amirkabir University of Technology, Tehran, Iran*

ABSTRACT: A complex network approach on a rough fracture is developed. In this manner, some hidden metric spaces (similarity measurements) between apertures profiles are set up and a general evolutionary network in two directions (in parallel and perpendicular to the shear direction) is constructed. Also, an algorithm (**CO**mplex **N**etworks on **A**pertures: CONA) is proposed in which evolving of a network is accomplished using preferential detachments and attachments of edges (based on a competition and game manner) while the number of nodes is fixed. Also, evolving of clustering coefficients and number of edges display similar patterns as well as are appeared in shear stress, hydraulic conductivity and dilation changes, which can be engaged to estimate shear strength distribution of asperities.

## 1 INTRODUCTION

Understanding of rock joint behaviors, either in single or mass form, under the several natural or artificial forces has been the subject of numerous researches during the evolution of rock mechanics field. Rock joint performance as a result of collective behavior of constructed elements (say fraction/pixel in each surface), interacting with each other, determines nonlinear picture of a changeable system. Obviously, one cannot predict the rich behavior of the whole by merely extrapolating from the treatment of its units (Boccara 2004, Vicsek 2002, Hakan 1989). Absent of fully prediction and nonlinear essences of behavior are prevalent gesture of complex systems. The term complex system formally refers to a system of many parts which are coupled in a nonlinear fashion.

When there are many non-linearities in a system (many components), behavior can be highly unpredictable. Complex systems research studies such behavior. Complex systems research overlaps with nonlinear dynamics research, but complex systems consist of a large number of mutually interacting dynamical parts. The success in describing of interwoven systems using physical tools as a major reductionism is associated with the simplifications of the interactions between the elements so that complexity reduction is a rescue pathway to regulation of approximated analyses of collective particles having swing states, complicated structures, and diversity of relations among elements. Complex networks have been developed in the several fields of science and engineering for example social, information, technological, biological and earthquake networks are the main distinguished networks (Albert & Barabasi 2002, Abe & Suzuki 2006).On the other hand, to catch on Hydro-mechanical and mechanical behavior of a rock joint, domination on to the surface morphology and its evolution as well as aperture is irrefutable. In addition to these procedures, the mechanical properties and hydraulic conductivity of the being joint are compared with the network properties. Upon this comparison, the distribution of shear strength –for each profile- is estimated.

## 2 COMPLEX NETWORKS

A network (graph) consists of nodes (sites-vertices) and edges (links- connections) connecting those (Wilson, 1996). If edges have their directions, the network is called directed otherwise network is undirected. One of the effective ways in visualization of a network is employing of an "adjacency matrix", which the matrix elements are 1(connection) or 0 (non-connection).

To set up a network on a single rock joint a network is assembled on the enclosed interval among two surfaces, i.e. complex aperture networks. To make edge between two nodes, a relation should be defined. In this study we assume that there are some hidden metric spaces between two nodes. In simple shape, the similarity between nodes is investigated while Euclidean distance is employed which is given as below:

$$d_{Euc.} = \sqrt{\sum_{p,q=1:n_p} (p(x_1, x_2, ..., x_n) - q(x_1, x_2, ..., x_n))^2} \qquad (1)$$

where $p$ and $q$ are the $i^{th}$ profiles. When $d \leq \xi$ an edge among two nodes is created. It can be proved the emerged network upon the mentioned way is an undirected network. The threshold $\xi$ depicts error level, usually can be assumed as 5-30 percent of maximum $d$. Let us introduce some properties of the undirected networks: clustering coefficient (C) and the degree distribution ($P(k)$). The clustering coefficient – or transitivity (Newman, 2003) - describes the degree to which k neighbors of a particular node are connected to each other. Our mean about neighbors is the connected nodes to the particular node. To better understand this concept and in the language of social networks the question "are my friends also friends to each other?" can be used. In fact clustering coefficient shows the collaboration (or synchronization and tendency) between the connected nodes to one. In terms of network topology, transitivity means the presence of heightened number of triangles in the network. The clustering coefficient measures the density of triangles in a network. Assume the $i^{th}$ node to have $k_i$ neighboring nodes. There can exist at most $k_i(k_i-1)/2$ edges between the neighbors (local complete graph). Define $c_i$ as the ratio

$$c_i = \frac{actual\ number\ of\ edges\ between\ the\ neighbors\ of\ the\ i^{th}\ node}{k_i(k_i-1)/2} \qquad (2)$$

Then, the clustering coefficient is given by the average of $c_i$ over all the nodes in the network:

$$C = \frac{1}{N}\sum_{i=1}^{N} c_i. \qquad (3)$$

For $k_i \leq 1$ we define $C \equiv 0$. The closer $C$ is to one the larger is the interconnectedness of the network. Consider that the clustering coefficient depicts the appeared triangles around each node. The connectivity distribution (or degree distribution), $P(k)$ is the probability of finding nodes with k edges in a network. In large networks, there will always be some fluctuations in the degree distribution. The large fluctuations from the average value ($<k>$) refers to the highly heterogeneous networks while homogeneous networks display low fluctuations.

## 3 RESULTS

In this part, we focus on the experimental results and mapping them into the complex networks. Our aim is underlined to find out the possible relations between the constructed networks properties and the current mechanical / hydro-mechanical properties of a rock joint which is un-der a constant normal stress and successive Shear Displacements (*SD*). The rock material was granite with the weight of 25.9 KN/m3 and uniaxial compressive strength of 172 MPa. An artificial rock joint was made at mid height of the specimen by using special joint creating apparatus, which has two horizontal jacks and a vertical jack (Sharifzadeh 2004, 2005). The final size of artificial joint is 180 mm in length, 100 mm in width and 80 mm in height. Using special mechanical units several parameters of this sample were measured. A virtual mesh having a square element size of 0.2 mm spread on each surface and each position height was measured by the laser scanner (Figure 1). Also a special hydraulic testing unit is employed to allow linear flow experiments (parallel to shear direction) while the rock joint is undergoing normal or shear loading. The details of the procedure and apparatus can be found in (Ghaffari et al. in press).

In this study, we consider only the evolution of apertures under constant normal stress and regular translational shear in which the lower surface has fixed position and upper one is displaced. By employing a suitable threshold value in Euclidian distance (Eq. (1)) and setting up a predestinated complex network on the X-profiles, gradual changes of the adjacency matrix form of the appeared networks can be inferred (Figure 2). Figure2 demonstrates after a phase transition step the similarities patterns are constrained to the adjacency of each profile. In other word, the decreasing of similarity between profiles coincides with the increasing of dilation.

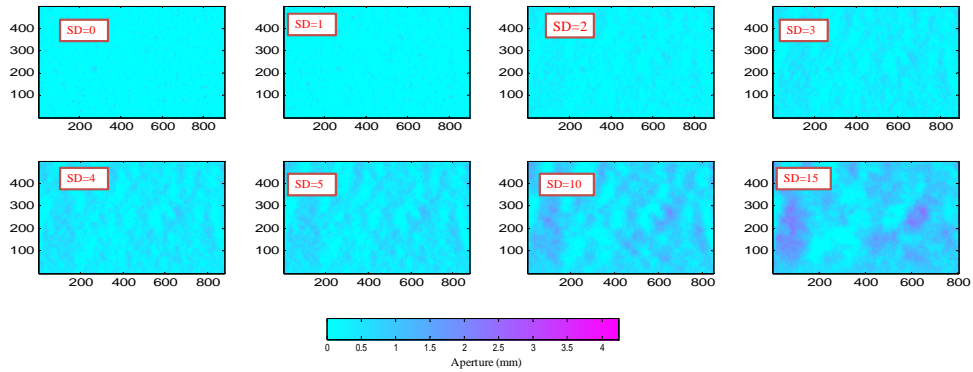

Figure 1. The evolution of apertures 'patterns under successive shear displacements and 3 Mpa normal stresses (the axis shows number of elements with a square element size of 0.2 mm)

This means that, after interlocking step (SD=1mm) the initial similarities patterns rapidly fall down in a constrained shape, i.e., the competition between making and destroying of edges are limited around a variable neighborhood radius of each node (profile). The neighborhood radius –in the final stages of disruption- changes from 2 to 20 pixels (0.4-4 mm), except for boundary profiles. The concentration of similarities takes place around 5-10 pixels where lower and higher values put in an almost symmetric shape which disclose a Gaussian distribution.

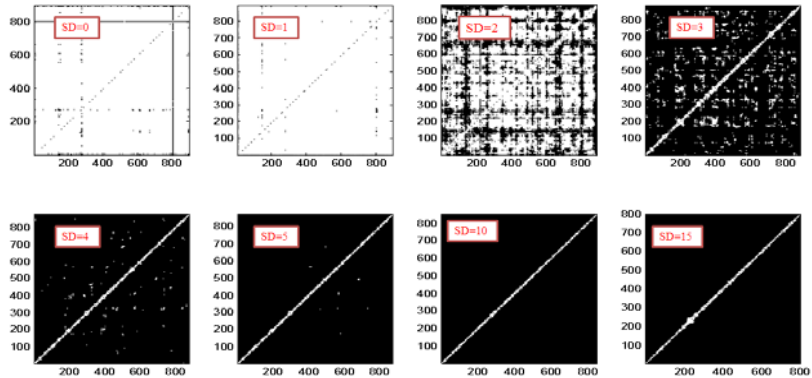

Figure 2. The evolution of X-profiles networks (adjacency matrix visualization) using Euclidean distance and $d \leq 5$ - (white occupations: connected elements)

At *X*-profiles, decreasing of the edges after *SD*=1 (Figure 2) coincides with a transition point (interval) which shows the system after an abrupt drop, goes towards a stable state. A similar behavior can be followed in the appropriate Y-profiles where the number of active nodes has a constant value and the variation of *K* doesn't show an intermediate step in declining procedure (discontinuous transition). In other word, the Y profiles exhibit a faster harmony with the forced shear loads.

Gradual changes of connectivity distributions, either in X or Y profiles, reveals a similar translational behavior: transformation from a nearly single value function to a semi-stable Gaussian distribution (Figure.3a, b). In parallel profiles with shear direction the neighborhood radius shows a lower interval rather than X-profiles. Also, in harmony with changes of edges, transition to a semi-stable stage occurs with more convergence rate. In other word, similarities of Y-profiles occur in higher constraints and readily (faster than opposite option). This event can be followed with more detail by employing a virtual state space which is built on the variations of $k_i$ and $c_i$ (Eq. (2)). Faster adapting in Y-profiles is the distinguished feature of $k_y^j - 1/c_y^j$ space rather than $k_x^i - 1/c_x^i$ space (Figure.4a, b) which there is a discerned fault among two main seeming clusters (at $SD = 2 \rightarrow SD = 4$).

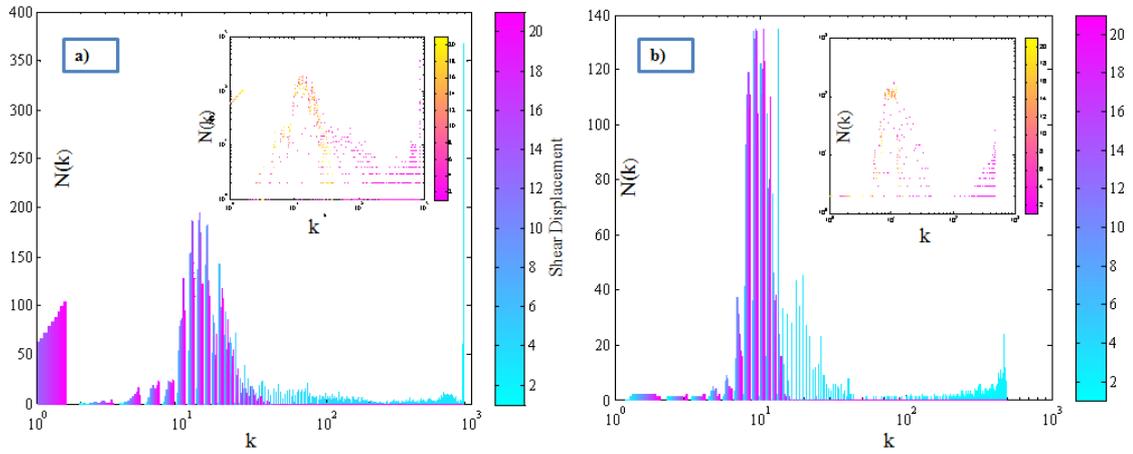

Figure 3. Frequency of nodes connectivity evolution over the shear displacements on: a) X-profiles and b) Y-profiles (Insets: results in log-log coordinate)

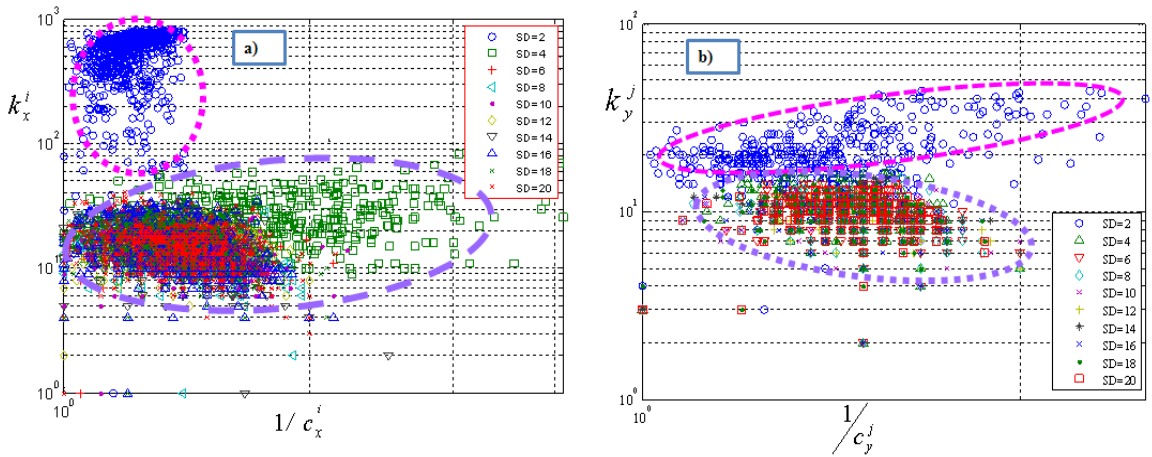

Figure 4. a) Evolving of $K_x^i - 1/c_x^i$ state space; b) $K_y^j - 1/c_y^j$ changes during shear displacements (Dashed ovals: the main seeming clusters)

The possible reason can be inferred from the lower resistance against transitional shear (and also easy leading of fluid flow). The complementary results using calculation of clustering coef-ficients emphasis that the inverse values of *C* during process ,in X-profiles, hand out like pattern as ones came out in the variations of shear stresses (Figure.5 a and c) while in the same manner and on Y-profiles resembles with the increment of hydraulic conductivity values (Figure.5 b and d). The coinciding of $1/C_Y$ with the changes of the hydraulic conductivity rather than the dilation behavior proves that the joint un-matching (or increment of mean aperture) is not only singular parameter in the fluid flow, as if reduction of joint roughness and the entire ensembles of the similarity behaviors of asperities are other possible agents.

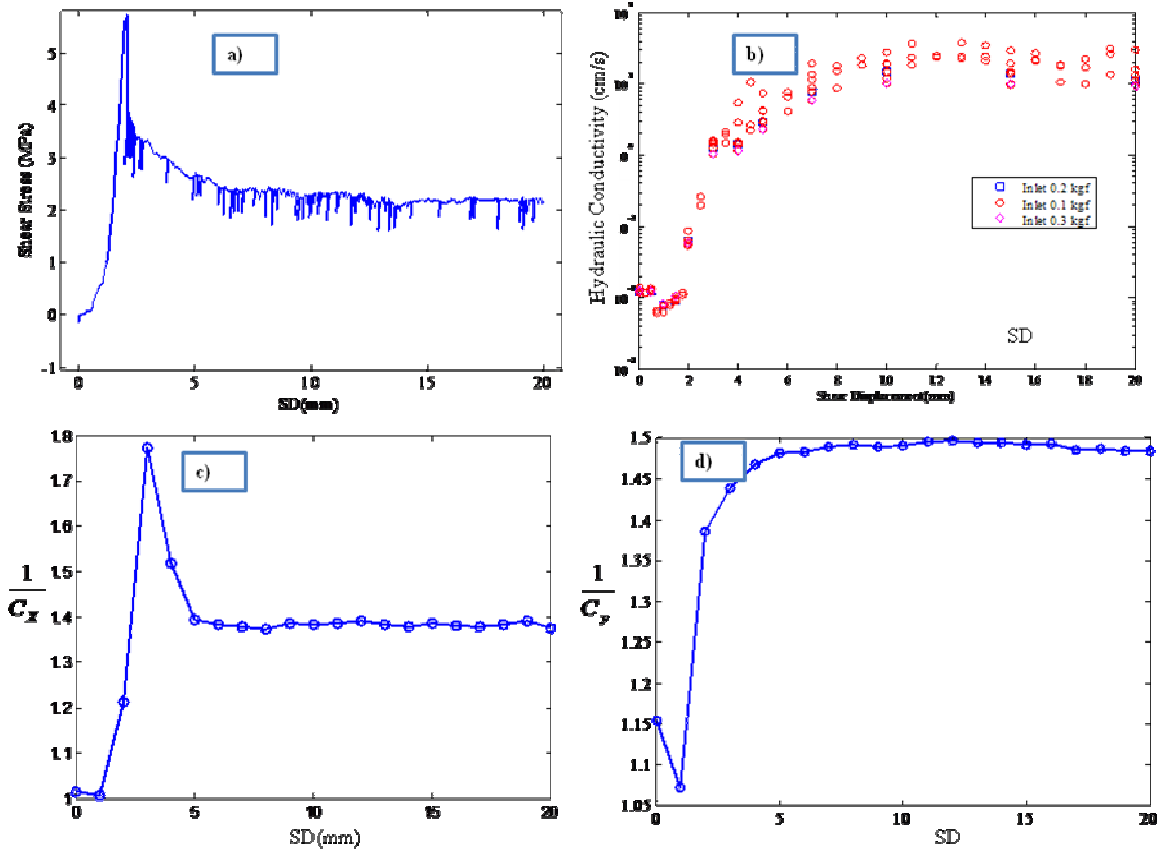

Figure 6.a) Shear Stress -Shear Displacement (Normal Stress: 3Mpa), b) Hydraulic Conductivity - Shear Displacement associated with 3 different cases, c) &d) Inverse of Clustering Coefficients -Shear Displacement on the X and Y profiles ,respectively.

Using an intelligent clustering method -based on competition in the hidden layer of neural network –self organizing feature map (SOM) (Kohonen,1987) - the appeared space $k_y^j - c_y^j$ with an elementary topology of $20 \times 20$ in second layer and over 500 epochs the dominant structures of this space are recognized ,are adapted with three main parts in Figure.7b: (1) increasing of $c_y^j$ -Decreasing $k_y^j$ ( increasing of $SD$ -Decreasing $K_h$ ) ,(2) increasing of $c_y^j$ -Growing $k_y^j$ ( increasing of $SD$ -Growing $K_h$ ) and (3) increasing of $c_y^j$ -slightly change of $k_y^j$ ( increasing of $SD$ - slightly change of $K_h$ ). The distribution of $c_x^i, c_y^j$ , based on the mentioned observations- can give a schematic view on the distribution of shear strength and hydraulic conductivity of each profile (Figure 8). Evolution of shear strengths frequency demonstrates that at initial steps the shear strength of elements cover a wide range while at meta-stable stages the concentration of strengths encases a Gaussian distribution with lesser scope. Alike procedure on $c_y^j$ -related with $K_h$ -reveals that at elementary stages of the shear displacement the overall pattern of distribution is formed and almost in other steps a significant disruption (transition) is not recognized (Figure 8.b).

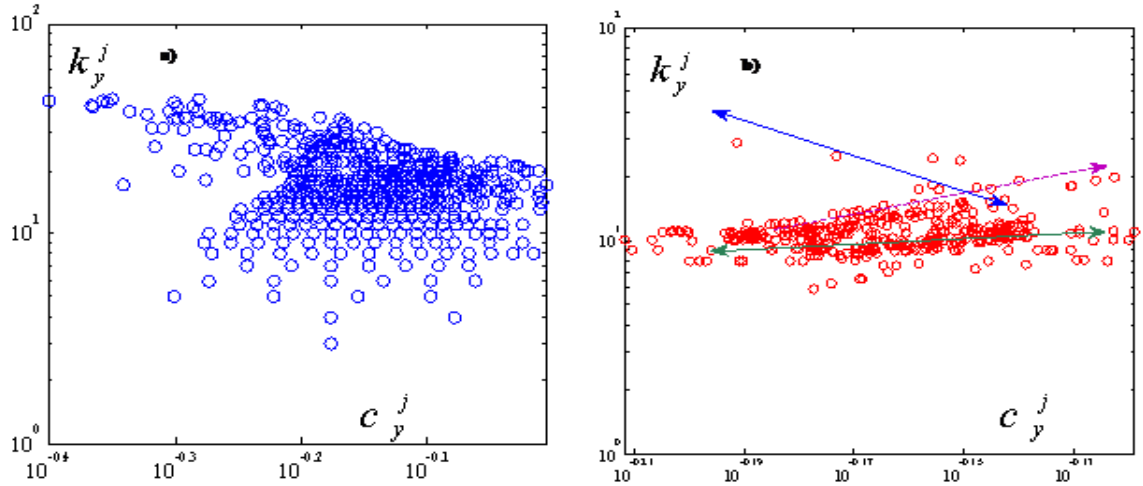

Figure 7. a) Evolving of $K_y^j - c_y^j$ state space; b) Clustering of $K_y^j - c_y^j$ space using SOM with initial topology $n_x = 20, n_y = 20$ in competition layer after 500 epochs training

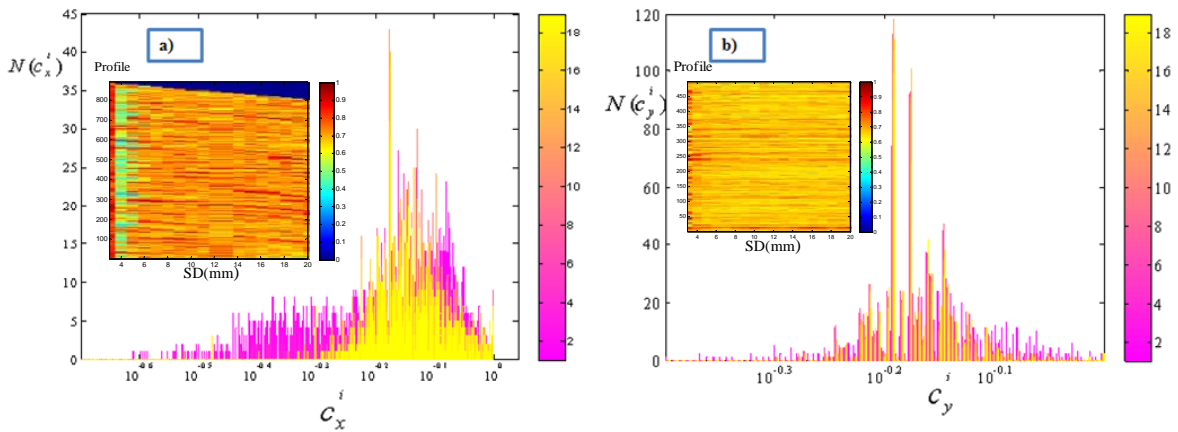

Figure 8. Evolution of frequency - $N(c_{x,y}^{i,j})$ - of: a) $c_x^i$ and b) $c_y^j$ over the shear displacements (insets show evolution of each profile)

Some other interesting results upon the employing of other measurements (metrics) on profiles can be ensued in (Ghaffari 2009).

## 4 AN ALGORITHM TO EVOLUTION OF COMPLEX APERTURE NETWORKS: **CONA**

In this part of this study we present a simple model which considers the edges attachments and detachments (decaying-growing of initial similarity patterns) of the network. Focusing on X-profiles indicates that after interlocking of asperities, the network edges fall down in a semi-stable occasion and slightly fluctuate, due to coherently construction –destruction procedure of edges while the appeared triangles around same nodes show high oscillations. This observation in the complex networks language can be interpreted as a continuously competition between attaching and detaching of edges while the nodes are fixed (Y-profiles) or decayed (X-profiles). In first steps (except inter locking step) the demolishing of edges are winner while in residual steps a game between two strategies is played.

On this construal, any proposed algorithm must consider either similar procedure in edges and intra-structures of profiles and distribution changes of the networks parameters. Here, we assume the decaying and local growing of networks are related to the random and preferential tactics. The meaning of preferential is related to the fitness of each node to absorb or repulse of edges. For simplicity, our model doesn't take in to account the nodes decaying (as it comes out in the X-profiles evolving) and attempt to capture the nearly possible mechanism(s) that govern the evolution of network topology. The steps of main algorithm (CONA) are as follows:

1) Start with $N$ nodes –which are fully connected to each other except in some elements.
2) At each time step, select $m_2$ nodes uniformly and attach the edges which their end points (node $i$ with the $k_i$ links ), are selected with a preferential probability ,is given as:

$$\Pi(A) = 1 - \frac{\beta k_i}{\sum k_i} \qquad (4)$$

which $\beta$ is tolerance parameter as with increasing, the attaching of connections is accomplished hardly.

3) At each time step select $m_1$ nodes uniformly and detach the edges which their end points (node $i$ with the $k_i$ links ), are selected with a preferential probability ,is given as:

$$\Pi(D) = \frac{\alpha k_i}{\sum k_i} \qquad (5)$$

$\alpha$ shows intolerance parameter and with increasing, the absorbing new ties will be decreased. One may find out a relation of tolerance- intolerance parameters with the internal and external characters of a given system such joint asperities, loads, etc. $m_1$ and $m_2$ are coefficients that indicate the rate of mass attachments (sprawl) or detachments of edges (genocide). From the point view of competition in social systems, above relations in selection of nodes (terminal nodes) can be interpreted as absorption of an active node to the "poor" individuals, in other word promoting of poor societies with adding new links, while other strategy tries to cut edges from those individuals who have expander relations. Let us state with a simple term:"The Poor are richer and the richer lose their possessions". Another strategy can be considered instead of (4), (5) is "The richer are richer –The poor are poorer".

In fact, the competition on the inducing of improving or worsening is an underlined key in this game. The effect of $\alpha, \beta$ variations over 1000 time steps and using N=40 (based on the initial random pattern -instead of fully connected network) has been shown in Figure.9. For a wide range of $\alpha, \beta$, the appeared patterns only show unstable state, however the transition from an unstable to stable state is occurred at a narrow range of resistant parameters (Figure.9). For $\alpha \geq 0.01, \beta \geq 28$ system loses most connections (collapsing) while for $\alpha \leq 0.01, \beta \leq 28$ most of nodes are connected to each other. Over $\alpha \leq 0.01, \beta \geq 28$, the system preserves the initial random pattern and doesn't change.

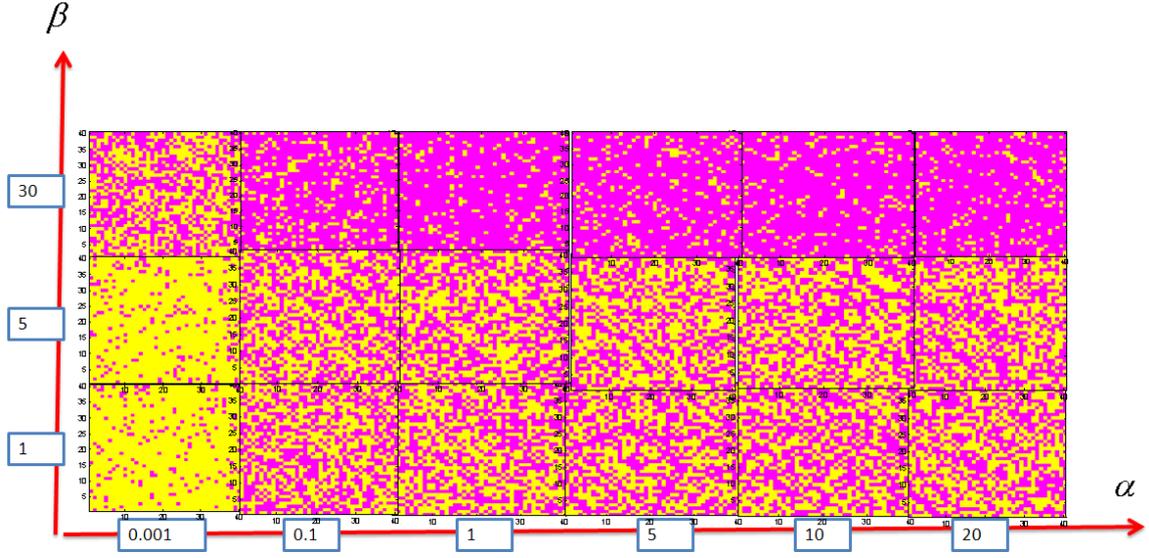

Figure9. Emerged patterns by variations of $\alpha, \beta$ in main CONA algorithm and

$$\Pi(A) = 1 - \frac{\beta k_i}{\sum k_i}, \Pi(D) = \frac{\alpha k_i}{\sum k_i}, m_1 = m_2 = 1$$

As it can be seen, CONA selects the initial and terminal nodes from the all of included individuals in the system when in the real provided features of apertures changes (Figure.2); the competition is constrained around the neighbors of each node. To reach this aim, we modify the main algorithm of CONA to the time-dependency growth form. In this way, the selection of terminal nodes-in growing step- from an initial fixed interval, i.e., $A = [1, N]$ is changed to $A' = [1 + \delta n(t), N - \delta n(t)]$ where $\delta n(t)$ is the rate of isolation and localization of the initial random node to find new relations. The final activity radius of initial node is terminated at: $A^{Final} = [n_{random} - \Delta, n_{random} + \Delta]$ where $\Delta$ is the final competition radius. Same as this manner, the $m_1$ and $m_2$ values can be regulated adaptively, i.e. over $0 \leq SD \leq 2$ (from starting point to interlocking step) attaching of edges is a prevalent procedure (so, $m_2 > m_1$). It seems at interlocking step (and its neighbor) $m_1 = m_2$ and gradually $m_1 > m_2$. With assuming $\delta n(t) = \lfloor \exp(t) \rfloor$, N=30, $\alpha = 1, \beta = 10, \Delta = 3$, the results of the modified CONA with $\{m_1 = 5, m_2 = 3\}$ and $\{m_1 = 10, m_2 = 3\}$ has been depicted (over 500 generations) in Figure.10 (a),(b), respectively.

One can follow the falling down rate of *K* and *C* and their relations with the changes of edges frequency: in high values of $m_1$ the intermediate steps are slightly ignored and an upward tendency to the final Gaussian distribution can be pursued (compare with Figure.2 and Figure.3). In another view, the inherent tolerance values can be attributed to the nodes as a specific internal property and a parameter relating with the initial anisotropy of elements appropriate with the captivity or liberation of edges. Behind this modification, one can carry on a game strategy, based on a synchronized strategy change on the complex network which is tied with evolutionary game theory and complex networks (Nowak 2006).

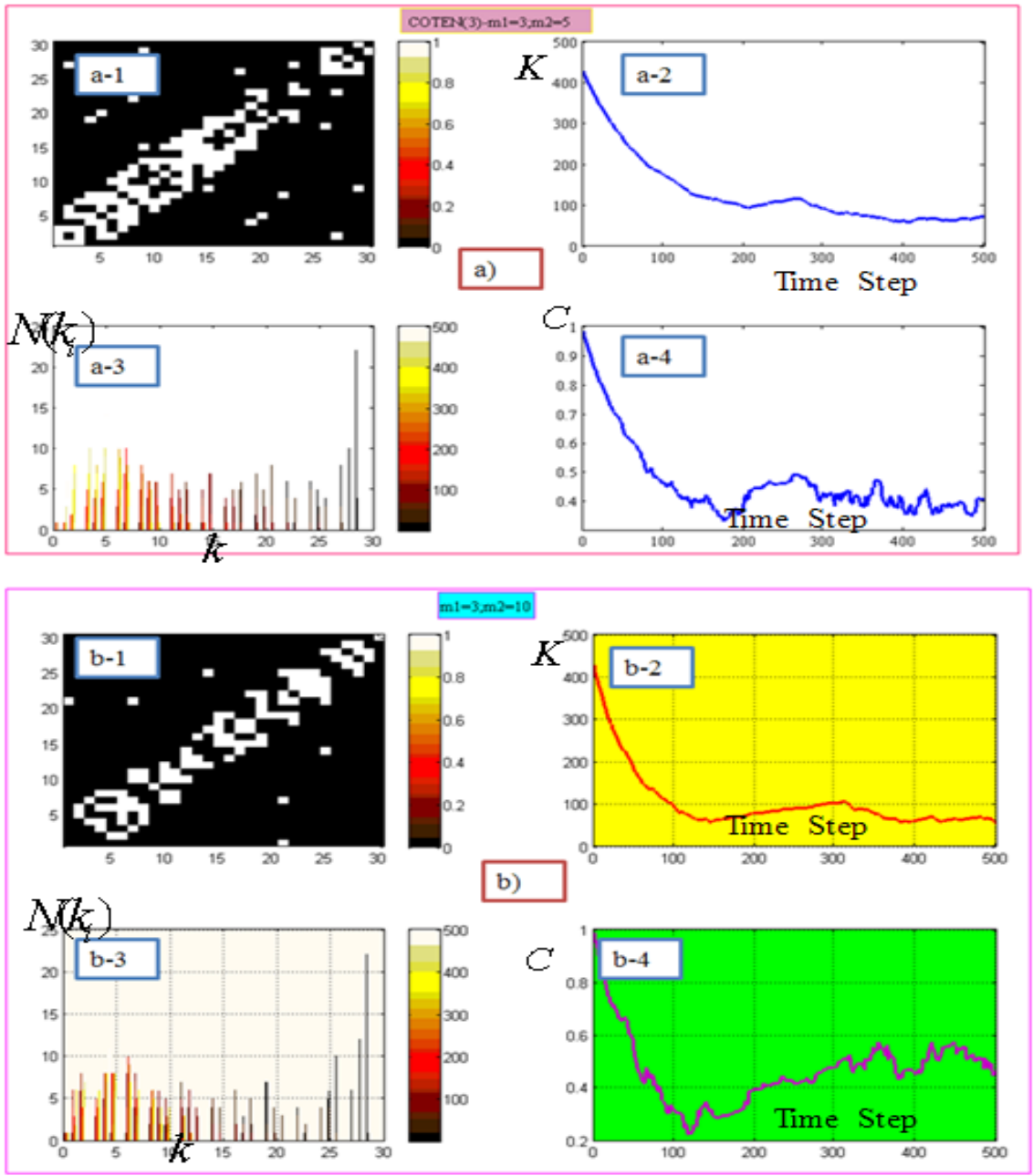

Figure10.The results of the modified CONA algorithm: a) $m_1 = 5, m_2 = 3$ and $\alpha = 1, \beta = 10$ -$a_1$: final adjacency matrix over 500 generations, $a_2$: K (number of total edges)-Time step, $a_3$: frequency of edges during procedure and $a_4$: C (Clustering Coefficient)-Time Step; b) $m_1 = 10, m_2 = 3$

## 5 CONCLUSION

The success in describing of interwoven systems using physical tools as a major reductionism is associated with the simplifications of the interactions between the elements where there is no possible vagueness. Employing of statistical mechanics tools gives a good framework for analysis of these systems. Also, possible relations between the building blocks of complex systems can be revealed in complex networks, which describe a wide range of our world systems. From this perspective and by considering of the complicated behavior of a rough fracture which was under the shear stress, a network associated with a popular similarity measure was designated at two separated directs of the shearing. The networks properties shed light a suitable coordination with the empirically obtained mechanical and hydro- mechanical characters of the being joint. Recognition of phase transition step, reforming of inter-structures evolutions, the weight of perpendicular profiles against resistance and procuring of great order in parallel profiles are some of the benefits of the captured networks. Thus, based on the real observations of the complex network changes, an algorithm (COmplex Networks on Apertures: CONA) was proposed in which evolving of a network is accomplished using preferential detachments and attachments of edges (based on a competition and game manner) while the number of nodes is fixed. Also, evolving of clustering coefficients and number of edges display similar patterns as well as are appeared in shear stress, hydraulic conductivity and dilation changes, which can be engaged to estimate shear strength distribution of asperities.